\documentclass[reprint,twocolumn]{revtex4}%
\usepackage{amsfonts}
\usepackage{amsmath}
\usepackage{amssymb}
\usepackage{charter}
\usepackage{graphicx}%
\providecommand{\U}[1]{\protect \rule{.1in}{.1in}}
\begin{document}
\title{Improving entanglement of even entangled coherent states by a coherent
superposition of photon subtraction and addition}
\author{{\small {Shi-You Liu,}}$^{1}${\small {Jie-Hui Huang,}}$^{1}${\small {Li-Yun
Hu}}$^{1,2}$\thanks{Corresponding author. E-mail: hlyun@jxnu.edu.cn.}%
,{\small {\  \ Zheng-Lu Duan}}$^{1}${\small {, Xue-Xiang Xu and Ying-Hua Ji}%
}$^{1,2}$}
\affiliation{$^{1}${\small Center for Quantum Science and Technology, Jiangxi Normal
University, Nanchang 330022, China}}
\affiliation{$^{2}${\small Key Laboratory of Optoelectronic and Telecommunication, Jiangxi
Normal University, Nanchang 330022}}

\begin{abstract}
{\small A new entangled quantum state is introduced by applying local coherent
superposition (}$r${\small $a^{\dag}+ta$) of photon subtraction and addition
to each mode of even entangled coherent state (EECS) and the properties of
entanglement are investigated. It is found that the Shchukin-Vogel
inseparability, the degree of entanglement and the average fidelity of quantum
teleportation of the EECS can be improved due to the coherent superposition
operation. The effects of improvement by coherent superposition operation are
better than those by single (}$a^{\dagger}${\small ) and two-photon
(}$a^{\dagger}b^{\dagger}${\small ) addition operations under a small region
of amplitude.}



\end{abstract}
\maketitle

\section{Introduction}

Quantum entanglement, as an important resource, plays a critical role in
quantum computation and quantum information processing \cite{1}. The two-mode
squeezed vacuum state exhibits quantum entanglement between the idle mode and
the signal mode, and is often used to as an entangled resource. However, the
source of practically available is characteristics of an finite degree of
entanglement. In order to realize the long-distance communication schemes of
quantum information, entanglement resources with higher degree of entanglement
are usually required. Thus it is not trivial to enhance the entanglement of
given entangled states.

For this purpose, many theoretical and experimental protocols are proposed to
generate such entangled states, which are plausible ways to obtain an
entangled state with higher entanglement degree by non-Gaussian operations,
such as adding or subtracting photons to/from quantum states \cite{2}. For
instance, the entanglement between Gaussian entangled states can be increased
by subtraction of single photon from Gaussian quadrature-entangled light
\cite{3,4}. The fidelity of quantum teleportation may be enhanced by
subtracting one-photon from each mode rather than only one mode \cite{5,6,7}.
In addition, the combination of photon-subtraction and photon-addition
operation has been used to generate nonclassical states. For example,
nonclassical states of optical field generated by photon
subtraction-then-addition or addition-then-subtraction operation \cite{8}.
Then multi-photons subtraction-then-addition or addition-then-subtraction is
further operated on coherent states. The nonclassicality of these states are
discussed according to photon-number distribution, Mandel Q parameter and
Wigner function \cite{9}. On the other hand, it is interesting to note that
coherent superposition of non-Gaussian operations is also proposed to
manipulate and enhance the entanglement of quantum states \cite{10,11}. In
Ref.\cite{10}, coherent superposition of photon subtraction and addition,
i.e., the $ta+ra^{\dagger}$ operator is used to transform a classical state to
a nonclassical one. Then in Ref. \cite{11} the coherent superposition is
further performed on two-mode squeezed vacuum for enhancing quantum
entanglement or non-Gaussian entanglement distillation. Recently, the
entanglement of a non-Gaussian entangled state is investigated by local photon
subtraction operation \cite{11a}. Thus these coherent superposition operations
can be seen as a useful tool in quantum information processing.

In this paper, we will introduce a kind of entangled state, coherent
superposition (CS) even entangled coherent state (EECS) $\left \vert \Psi
_{+}\left(  \alpha,0\right)  \right \rangle $ (see below Eq.(\ref{ht1})),
generated by applying coherent superposition operation $\left(  t_{A}%
a+r_{A}a^{\dagger}\right)  ^{m}\left(  t_{B}b+r_{B}b^{\dagger}\right)  ^{n}$
to each local mode of two-mode EECSs and investigate the properties of
entanglement. The photon-subtraction ($a^{m}b^{n}$,$\ r_{A}=r_{B}=0$) EECS and
the photon-addition ($a^{\dagger m}b^{\dagger n}$, $r_{A}=r_{B}=1$) EECS can
be considered as two special cases of the CS-EECSs. Practically speaking, we
investigate the degree of entanglement by using the concurrence, the
inseparability by using Shchukin-Vogel (SV) criteria, and the average fidelity
of quantum teleportation to explore how these properties can be enhanced by
the CS operation. It is shown that all these entanglement characteristics can
be improved by operating the CS operation on the EECSs under a certain
amplitude region. Sa far we know, there is no report about this issue in the
literature to date.

This paper is arranged as follows. In section 2, we shall introduce the
CS-EECSs, and derive the normalization factor by using the generating function
of single-variable Hermite polynomials, which is necessary for further
discussing the entanglement properties of quantum states. It is shown that the
factor is just related to Hermite polynomials. The improvement of entanglement
measured by Shchukin-Vogel criteria and the concurrence in sections 3 and 4,
respectively. Section 5 is devoted to studying the average fidelity of quantum
teleportation with continuous variable by using the CS-EECSs as the entangled
channel. It is found that the effective teleportation (fidelity%
$>$%
1/2) can be enhanced by CS operation on the EECSs in a small amplitude region.
The main results are involved in the last section.

\section{Two-mode coherent superposition EECSs}

In this section, we introduce the two-mode CS-EECSs. First, let us begin with
the EECSs \cite{12}, which is defined by%
\begin{equation}
\left \vert \Psi_{+}\left(  \alpha,0\right)  \right \rangle =N_{+,\alpha}\left(
\left \vert \alpha,\alpha \right \rangle +\left \vert -\alpha,-\alpha \right \rangle
\right)  , \label{ht1}%
\end{equation}
where $\left \vert \alpha,\alpha \right \rangle \equiv \left \vert \alpha
\right \rangle _{a}\otimes \left \vert \alpha \right \rangle _{b}$ with $\left \vert
\alpha \right \rangle _{a}$ and $\left \vert \alpha \right \rangle _{b}$ being the
usual coherent states in $a$ and $b$\ modes, respectively, and $\left(
N_{+}\left(  \alpha,0\right)  \right)  ^{-2}=2[1+e^{-4\left \vert
\alpha \right \vert ^{2}}]$ is the normalization constants. By performing
repeatedly the two-mode CS operations $\mathfrak{A}_{m}\equiv \left(
r_{A}a^{\dagger}+t_{A}a\right)  ^{m}$ and $\mathfrak{B}_{n}\equiv \left(
r_{B}b^{\dagger}+t_{B}b\right)  ^{n}$ on the ECSs, one can obtain the output
state as%
\begin{equation}
\left \vert \Psi_{+}\left(  \alpha,m,n\right)  \right \rangle =N_{+,m,n}%
\mathfrak{A}_{m}\mathfrak{B}_{n}\left(  \left \vert \alpha,\alpha \right \rangle
+\left \vert -\alpha,-\alpha \right \rangle \right)  , \label{ht2}%
\end{equation}
where $N_{+,m,n}$ represents the normalization factor and $\left \vert
t_{j}\right \vert ^{2}+\left \vert r_{j}\right \vert ^{2}=1$ ($j=A,B$). In
particular, when $t_{A}=t_{B}=0$ and $r_{A}=r_{B}=1,$ Eq.(\ref{ht2}) just
reduces to the two-mode excited EECSs \cite{13,13a}, $\left \vert \Psi
_{+}\left(  \alpha,m,n\right)  \right \rangle =N_{+,m,n}a^{\dagger m}b^{\dagger
n}\left(  \left \vert \alpha,\alpha \right \rangle +\left \vert -\alpha
,-\alpha \right \rangle \right)  $. The properties of entanglement of
single-mode excited EECSs (say $n=0$) is discussed in details \cite{14}.

Next, we derive the normalization factor $N_{+,m,n}$, which is important for
further discussing the properties of statistics and entanglement for quantum
states. For this purpose, using the formula $e^{A+B}=e^{A}e^{B}e^{-\left[
A,B\right]  /2}=e^{B}e^{A}e^{\left[  A,B\right]  /2},$ which is valid for
$[A,\left[  A,B\right]  ]=[B,\left[  A,B\right]  ]=0,$ one can find, for
instance,
\begin{align}
&  \left(  r_{A}a^{\dagger}+t_{A}a\right)  ^{m}\left \vert \alpha \right \rangle
\nonumber \\
&  =\left.  e^{-\frac{1}{2}\left \vert \alpha \right \vert ^{2}}\frac
{\partial^{m}}{\partial s^{m}}e^{\frac{1}{2}s^{2}t_{A}r_{A}+\alpha st_{A}%
}e^{\left(  sr_{A}+\alpha \right)  a^{\dagger}}\left \vert 0\right \rangle
\right \vert _{s=0}, \label{ht3}%
\end{align}
which leads to (for simplicity, $t_{i}$ and $r_{i}$ are considered as real)%
\begin{align}
A_{1}  &  \equiv \left \langle \alpha \right \vert \left(  r_{A}a+t_{A}a^{\dagger
}\right)  ^{m}\left(  r_{A}a^{\dagger}+t_{A}a\right)  ^{m}\left \vert
\alpha \right \rangle \nonumber \\
&  =e^{-\left \vert \alpha \right \vert ^{2}}\frac{\partial^{2m}}{\partial
s^{m}\partial \tau^{m}}e^{\frac{1}{2}\left(  s^{2}+\tau^{2}\right)  t_{A}%
r_{A}+\alpha st_{A}+\alpha^{\ast}\tau t_{A}}\nonumber \\
&  \times \left.  \left \langle 0\right \vert e^{\left(  \tau r_{A}+\alpha^{\ast
}\right)  a}e^{\left(  sr_{A}+\alpha \right)  a^{\dagger}}\left \vert
0\right \rangle \right \vert _{s,\tau=0}\nonumber \\
&  =\frac{\partial^{2m}}{\partial s^{m}\partial \tau^{m}}\exp \left \{
\frac{t_{A}r_{A}}{2}\left(  s^{2}+\tau^{2}\right)  \right \} \nonumber \\
&  \times \left.  \exp \left(  \tau sr_{A}^{2}+s\left(  t_{A}\alpha+r_{A}%
\alpha^{\ast}\right)  +\tau \left(  t_{A}\alpha^{\ast}+r_{A}\alpha \right)
\right)  \right \vert _{s,\tau=0}. \label{ht4}%
\end{align}
Using the following formula \cite{15}%
\begin{align}
&  \frac{\partial^{2m}}{\partial t^{m}\partial s^{m}}\exp \left[  R_{1}%
s+R_{1}^{\ast}t+R_{3}ts+R_{4}\left(  t^{2}+s^{2}\right)  \right]
_{t=s=0}\nonumber \\
&  =\sum_{l=0}^{m}\frac{R_{3}^{l}R_{4}^{m-l}\left(  m!\right)  ^{2}}{l!\left[
\left(  m-l\right)  !\right]  ^{2}}\left \vert H_{m-l}[R_{1}/(2i\sqrt{R_{4}%
})]\right \vert ^{2}, \label{ht5}%
\end{align}
where $H_{m}\left(  x\right)  $ is the single-variable Hermite polynomials,
one can put Eq.(\ref{ht4}) into the following form%
\begin{equation}
A_{1}=\sum_{l=0}^{m}\frac{\left(  m!\right)  ^{2}t_{A}^{m-l}r_{A}^{m+l}%
}{2^{m-l}l!\left[  \left(  m-l\right)  !\right]  ^{2}}\left \vert H_{m-l}%
[\bar{\alpha}_{A}]\right \vert ^{2}. \label{ht6}%
\end{equation}
Here we have set $\bar{\alpha}_{A}=\left(  t_{A}\alpha+r_{A}\alpha^{\ast
}\right)  /(\mathtt{i}\sqrt{2t_{A}r_{A}})$. In a similar way, one can derive%
\begin{align}
A_{2}  &  \equiv \left \langle -\alpha \right \vert \left(  r_{A}a+t_{A}%
a^{\dagger}\right)  ^{m}\left(  r_{A}a^{\dagger}+t_{A}a\right)  ^{m}\left \vert
\alpha \right \rangle \nonumber \\
&  =e^{-2\left \vert \alpha \right \vert ^{2}}\sum_{l=0}^{m}\frac{(-1)^{m-l}%
\left(  m!\right)  ^{2}t_{A}^{m-l}r_{A}^{m+l}}{2^{m-l}l!\left[  \left(
m-l\right)  !\right]  ^{2}}\left \vert H_{m-l}[\bar{\beta}_{A}]\right \vert
^{2}, \label{ht7}%
\end{align}
and%
\begin{align}
B_{1}  &  \equiv \left \langle \alpha \right \vert \left(  r_{B}b+t_{B}b^{\dagger
}\right)  ^{n}\left(  r_{B}b^{\dagger}+t_{B}b\right)  ^{n}\left \vert
\alpha \right \rangle \nonumber \\
&  =\sum_{l=0}^{n}\frac{\left(  n!\right)  ^{2}t_{B}^{n-l}r_{B}^{n+l}}%
{2^{n-l}l!\left[  \left(  n-l\right)  !\right]  ^{2}}\left \vert H_{n-l}%
[\bar{\alpha}_{B}]\right \vert ^{2}, \label{ht8}%
\end{align}
as well as%
\begin{align}
B_{2}  &  \equiv \left \langle -\alpha \right \vert \left(  r_{B}b+t_{B}%
b^{\dagger}\right)  ^{n}\left(  r_{B}b^{\dagger}+t_{B}b\right)  ^{n}\left \vert
\alpha \right \rangle \nonumber \\
&  =e^{-2\left \vert \alpha \right \vert ^{2}}\sum_{l=0}^{n}\frac{(-1)^{n-l}%
\left(  n!\right)  ^{2}t_{B}^{n-l}r_{B}^{n+l}}{2^{n-l}l!\left[  \left(
n-l\right)  !\right]  ^{2}}\left \vert H_{n-l}[\bar{\beta}_{B}]\right \vert
^{2}, \label{ht9}%
\end{align}
where $\bar{\beta}_{A}=\left(  r_{A}\alpha-\alpha^{\ast}t_{A}\right)
/(\mathtt{i}\sqrt{2t_{A}r_{A}})$, $\bar{\alpha}_{B}=\left(  t_{B}\alpha
+r_{B}\alpha^{\ast}\right)  /(\mathtt{i}\sqrt{2t_{B}r_{B}})$ and $\bar{\beta
}_{B}=\left(  r_{B}\alpha-\alpha^{\ast}t_{B}\right)  /(\mathtt{i}\sqrt
{2t_{B}r_{B}})$.

Thus the normalization factor $N_{+,m,n}$ can be given by%
\begin{equation}
\left(  N_{+,m,n}\right)  ^{-2}=2\left(  A_{1}B_{1}+A_{2}B_{2}\right)
.\label{ht10}%
\end{equation}
It is noted that when $m=n=0$, then $A_{1}=B_{1}=1,$ and $A_{2}=B_{2}%
=e^{-2\left \vert \alpha \right \vert ^{2}}$, thus Eq.(\ref{ht2}) becomes the
usual EECSs in Eq.(\ref{ht1}) with $N_{+,0,0}^{-2}=2[1+\exp(-4\left \vert
\alpha \right \vert ^{2})].$ For the case of $t_{A}=t_{B}=0$ and $r_{A}%
=r_{B}=1,$corresponding to the two-mode excited EECSs, from Eqs.(\ref{ht4}%
)-(\ref{ht9}), one can see that
\begin{equation}
A_{1}=m!L_{m}(-\left \vert \alpha \right \vert ^{2}),B_{1}=n!L_{n}(-\left \vert
\alpha \right \vert ^{2}),\label{ht12}%
\end{equation}
and
\begin{align}
A_{2} &  =m!\exp(-2\left \vert \alpha \right \vert ^{2})L_{m}(\left \vert
\alpha \right \vert ^{2}),\nonumber \\
B_{2} &  =n!\exp(-2\left \vert \alpha \right \vert ^{2})L_{n}(\left \vert
\alpha \right \vert ^{2}),\label{ht13}%
\end{align}
thus the normalization factor Eq.(\ref{ht10}) reduces to%
\begin{align}
\left(  N_{+,m,n}\right)  ^{-2}  & =2m!n![L_{m}(-\left \vert \alpha \right \vert
^{2})L_{n}(-\left \vert \alpha \right \vert ^{2})\nonumber \\
& +e^{-4\left \vert \alpha \right \vert ^{2}}L_{m}(\left \vert \alpha \right \vert
^{2})L_{n}(\left \vert \alpha \right \vert ^{2})],\label{ht14}%
\end{align}
where $L_{m}(x)$ is the m-order Laguerre polynomial. Eq.(\ref{ht14}) is just
the normalization factor of the two-mode excited EECSs.

\section{Shchukin-Vogel inseparability}

For a bipartite continuous-variable state, here we will take the
Shchukin-Vogel (SV) criteria \cite{20} to describe the inseparability of the
CS-EECSs. According to the SV criteria, the sufficient condition of
inseparability (entanglement) is given by \cite{20a}
\begin{equation}
S_{+}\equiv \left \langle a^{\dag}a-\frac{1}{2}\right \rangle \left \langle
b^{\dag}b-\frac{1}{2}\right \rangle -\left \langle a^{\dag}b^{\dag}\right \rangle
\left \langle ab\right \rangle <0. \label{SV1}%
\end{equation}
Thus we only need to calculate three average values $\left \langle a^{\dag
}a\right \rangle ,\left \langle b^{\dag}b\right \rangle ,\left \langle
ab\right \rangle $ to obtain $S_{+}$.

Here, we consider the property of entanglement of the CS-ECSs with $m=n=1$,
$\left \vert \Psi_{+}\left(  \alpha,1,1\right)  \right \rangle =N_{+,1,1}%
\mathfrak{A}_{1}\mathfrak{B}_{1}\left(  \left \vert \alpha,\alpha \right \rangle
+\left \vert -\alpha,-\alpha \right \rangle \right)  ,$ where $N_{+,1,1}%
^{-2}=2\left(  A_{1}B_{1}+A_{2}B_{2}\right)  $ and%
\begin{align}
A_{1} &  =\left \vert t_{A}\alpha+r_{A}\alpha^{\ast}\right \vert ^{2}+r_{A}%
^{2},\nonumber \\
B_{1} &  =\left \vert t_{B}\alpha+r_{B}\alpha^{\ast}\right \vert ^{2}+r_{B}%
^{2}\nonumber \\
A_{2} &  =\left(  r_{A}^{2}-\left \vert r_{A}\alpha-\alpha^{\ast}%
t_{A}\right \vert ^{2}\right)  e^{-2\left \vert \alpha \right \vert ^{2}%
},\nonumber \\
B_{2} &  =\left(  r_{B}^{2}-\left \vert r_{B}\alpha-\alpha^{\ast}%
t_{B}\right \vert ^{2}\right)  e^{-2\left \vert \alpha \right \vert ^{2}%
}.\label{SV2}%
\end{align}
Under the state $\left \vert \Psi_{+}\left(  \alpha,1,1\right)  \right \rangle
$, we can derive three average values as
\begin{align}
\left \langle a^{\dag}a\right \rangle _{1,1} &  =\frac{O_{A1}B_{1}+O_{A2}B_{2}%
}{A_{1}B_{1}+A_{2}B_{2}},\nonumber \\
\left \langle b^{\dag}b\right \rangle _{1,1} &  =\frac{O_{B1}A_{1}+O_{B2}A_{2}%
}{A_{1}B_{1}+A_{2}B_{2}},\nonumber \\
\left \langle ab\right \rangle _{1,1} &  =\frac{R_{A1}R_{B1}+R_{A2}R_{B2}}%
{A_{1}B_{1}+A_{2}B_{2}},\label{SV3}%
\end{align}
where $A_{1,2}$ and $B_{1,2}$\ are defined in Eqs.(\ref{SV2}) and we have set%
\begin{align}
O_{j1} &  =\left \vert \alpha \right \vert ^{4}+r_{j}^{2}(1+3\left \vert
\alpha \right \vert ^{2})\nonumber \\
&  +t_{j}r_{j}\left(  \alpha^{\ast2}+\alpha^{2}\right)  (1+\left \vert
\alpha \right \vert ^{2}),\nonumber \\
O_{j2} &  =e^{-2\left \vert \alpha \right \vert ^{2}}\left[  \left \vert
\alpha \right \vert ^{4}+r_{j}^{2}(1-3\left \vert \alpha \right \vert ^{2})\right.
\nonumber \\
&  \left.  +t_{j}r_{j}\left(  \alpha^{\ast2}+\alpha^{2}\right)  (1-\left \vert
\alpha \right \vert ^{2})\right]  ,\label{SV4}%
\end{align}
and%

\begin{align}
R_{j1}  &  =\left \vert \alpha \right \vert ^{2}\alpha+t_{j}r_{j}(\alpha
^{3}+\left \vert \alpha \right \vert ^{2}\alpha^{\ast}+\alpha^{\ast})+2r_{j}%
^{2}\alpha,\nonumber \\
R_{j2}  &  =\left[  -\left \vert \alpha \right \vert ^{2}\alpha+t_{j}r_{j}%
(\alpha^{3}+\left \vert \alpha \right \vert ^{2}\alpha^{\ast}-\alpha^{\ast
})+2r_{j}^{2}\alpha \right]  e^{-2\left \vert \alpha \right \vert ^{2}%
},\label{SV5}\\
&  \left(  j=A,B\right)  .\nonumber
\end{align}
Using Eqs.(\ref{SV2})-(\ref{SV5}) we can get the expression of $S_{+}$. Here
we appeal to the numerical calculation for further discussion. In Fig. 2, we
plot the sufficient condition of inseparability as the function of $\alpha$
(as a real number) for the EECSs and the CS-EECSs with several different
values of $t$. It is shown that the condition $S_{+}<0$ can be satisfied only
when $\alpha$ exceeds a certain threshold value ($\alpha>0.567$ about). In
fact, under the EECSs we have $\left \langle a^{\dag}a\right \rangle
=\left \langle b^{\dag}b\right \rangle =\left \vert \alpha \right \vert ^{2}%
\tanh(2\left \vert \alpha \right \vert ^{2}),$ and $\left \langle ab\right \rangle
=\alpha^{2},$ thus%
\begin{equation}
S_{+}=\left(  \left \vert \alpha \right \vert ^{2}\tanh(2\left \vert
\alpha \right \vert ^{2})-\frac{1}{2}\right)  ^{2}-\left \vert \alpha \right \vert
^{4}<0, \label{SV6}%
\end{equation}
which lead to the following condition:%
\begin{equation}
2\left \vert \alpha \right \vert ^{2}\left(  \tanh(2\left \vert \alpha \right \vert
^{2})+1\right)  >1. \label{SV7}%
\end{equation}
For the coherent superposition operation ($0<t<1$), to satisfy the SV
criteria, the value of $\alpha$ must be larger than a certain threshold value
which decreases with the increasing $t$ (For simplicity we take $t_{A,B}=t$
and $r_{A,B}=r$, $\alpha$ as real). From Fig.1, we can see that (i) these
threshold values in CS-EECSs are smaller than that in EECSs; (ii) the
inseparability in the CS-EECSs is stronger than in the EECSs and the
photon-added EECSs ($a^{\dag}b^{\dag}\left \vert \Psi_{+}\left(  \alpha
,0\right)  \right \rangle $, at $t=0$); This indicates that the coherent
operation on both local modes improves the inseparability better than
photon-addition operation $a^{\dag}b^{\dag}$ in the small amplitude region.
(iii) the photon-subtraction operation $ab$ (at $t=1$) can not be used to
enhance the inseparability of the EECSs $\left \vert \Psi_{+}\left(
\alpha,0\right)  \right \rangle $ due to the fact that $\left \vert \Psi
_{+}\left(  \alpha,0\right)  \right \rangle $ is the eigenstate of pair
annihilation operator $ab$. From the point of SV criteria, the inseparability
of the EECSs can be improved better by operating CS on both local modes than
photon-addition operation ($a^{\dag}b^{\dag}$)\ under a certain condition.

\begin{figure}[ptb]
\label{Fig0} \centering \includegraphics[width=7cm]{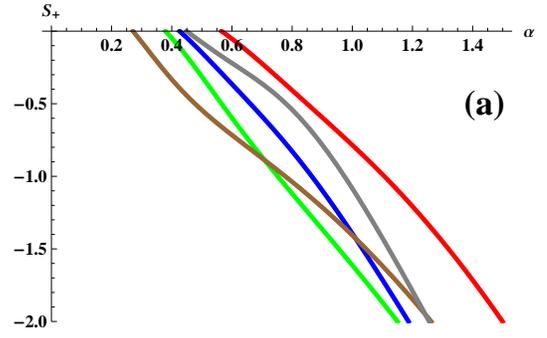}\caption{{}%
{\protect \small (Color online) The sufficient condition of inseparability as
the function of }$\alpha${\protect \small for the ECSs (red line) and CS-ECSs
with several different values of }$t${\protect \small : t=0 (Gray line), t=0.2
(Blue line), t=0.6 (Green line) and t=0.9 (Brown).}}%
\end{figure}

\section{Degree of entanglement of the CS-EECSs}

Next, we calculate the amount of entanglement of the CS-EECSs and investigate
the influence of the two-mode coherent superposition operation on the
entanglement of the CS-EECSs. For continuous-variables-type entangled states
like (\ref{ht2}), the degree of quantum entanglement of the bipartite
entangled states can be measured in terms of the concurrence. Following the
approach \cite{21,22,23}, for two systems involving nonorthogonal states
defined as
\begin{equation}
\left \vert \Psi \right \rangle =N\left[  \mu \left \vert \eta \right \rangle
_{a}\otimes \left \vert \gamma \right \rangle _{b}+\nu \left \vert \xi \right \rangle
_{a}\otimes \left \vert \delta \right \rangle _{b}\right]  , \label{ht27}%
\end{equation}
where $\left \vert \eta \right \rangle _{a},$ $\left \vert \xi \right \rangle _{a}$
and $\left \vert \gamma \right \rangle _{b}$, $\left \vert \delta \right \rangle
_{b}$ are normalized states of system $a$ and $b$, respectively, the
concurrence can be calculated as \cite{24,24a}%
\begin{equation}
C=\frac{2\left \vert \mu \right \vert \left \vert \nu \right \vert \left[
(1-\left \vert p_{1}\right \vert ^{2})(1-\left \vert p_{2}\right \vert
^{2})\right]  ^{1/2}}{\left \vert \mu \right \vert ^{2}+\left \vert \nu \right \vert
^{2}+2\text{Re}\left(  \mu^{\ast}\nu p_{1}p_{2}^{\ast}\right)  }, \label{ht28}%
\end{equation}
where the normalization constant is $N^{-2}=\left \vert \mu \right \vert
^{2}+\left \vert \nu \right \vert ^{2}+2$Re$\left(  \mu^{\ast}\nu p_{1}%
p_{2}^{\ast}\right)  ,$ with complex numbers $\mu$ and $\nu$, and
$p_{1}=\left.  _{a}\left \langle \eta|\xi \right \rangle _{a}\right.
,p_{2}=\left.  _{b}\left \langle \delta|\gamma \right \rangle _{b}\right.  $.
When $\mu=\nu,$ the concurrence $C$ becomes
\begin{equation}
C_{+}=\frac{\sqrt{(1-\left \vert p_{1}\right \vert ^{2})(1-\left \vert
p_{2}\right \vert ^{2})}}{1+\text{Re}\left(  p_{1}p_{2}^{\ast}\right)  },
\label{ht29}%
\end{equation}
which only dependent on the overlaps $\left \langle \eta|\xi \right \rangle $ and
$\left \langle \gamma|\delta \right \rangle $.

In order to obtain the amount of entanglement of the CS-ECSs, we introduce the
following normalized states:
\begin{align}
\left \vert \pm \alpha \right \rangle _{m} &  =A_{1}^{-1/2}\left(  r_{A}%
a^{\dagger}+t_{A}a\right)  ^{m}\left \vert \pm \alpha \right \rangle ,\nonumber \\
\left \vert \pm \alpha \right \rangle _{n} &  =B_{1}^{-1/2}\left(  r_{B}%
b^{\dagger}+t_{B}b\right)  ^{n}\left \vert \pm \alpha \right \rangle ,\label{ht30}%
\end{align}
then the CS-ECSs $\left \vert \Psi_{+}\left(  \alpha,m,n\right)  \right \rangle
$ in terms of four normalized states can be reformed as
\begin{align}
\left \vert \Psi_{+}\left(  \alpha,m,n\right)  \right \rangle  & =N_{+,m,n}%
\left(  A_{1}B_{1}\right)  ^{1/2}\nonumber \\
& \left(  \left \vert \alpha \right \rangle _{m}\otimes \left \vert \alpha
\right \rangle _{n}+\left \vert -\alpha \right \rangle _{m}\otimes \left \vert
-\alpha \right \rangle _{n}\right)  .\label{ht31}%
\end{align}
Thus one can find
\begin{align}
p_{1} &  =\left.  _{m}\left \langle \alpha \right.  \left \vert -\alpha
\right \rangle _{m}\right.  =\frac{A_{2}}{A_{1}},\nonumber \\
p_{2} &  =\left.  _{n}\left \langle -\alpha \right.  \left \vert \alpha
\right \rangle _{n}\right.  =\frac{B_{2}}{B_{1}}.\label{ht32}%
\end{align}
Substituting Eq.(\ref{ht32}) into Eq.(\ref{ht29}) one can get%
\begin{equation}
C_{+}=\frac{\sqrt{(A_{1}^{2}-A_{2}^{2})(B_{1}^{2}-B_{2}^{2})}}{A_{1}%
B_{1}+A_{2}B_{2}},\label{ht33}%
\end{equation}
where $A_{1},B_{1}$ and $A_{2},B_{2}$ are defined in Eqs.(\ref{ht6}%
)-(\ref{ht9}). Eq.(\ref{ht33}) is the analytical expression of concurrence for
the CS-EECSs.

In particular, for the two-mode excited EECSs, i.e., $t_{A}=t_{B}=0$ and
$r_{A}=r_{B}=1,$ leading to $A_{1}=L_{m}(-\left \vert \alpha \right \vert
^{2}),B_{1}=L_{n}(-\left \vert \alpha \right \vert ^{2})$ and $A_{2}%
=e^{-2\left \vert \alpha \right \vert ^{2}}L_{m}(\left \vert \alpha \right \vert
^{2})$, $B_{2}=e^{-2\left \vert \alpha \right \vert ^{2}}L_{n}(\left \vert
\alpha \right \vert ^{2}),$ then Eq.(\ref{ht33}) reduces to
\begin{align}
C_{+}  & =\frac{\sqrt{L_{m}^{2}(-\left \vert \alpha \right \vert ^{2}%
)-e^{-4\left \vert \alpha \right \vert ^{2}}L_{m}^{2}(\left \vert \alpha
\right \vert ^{2})}}{L_{m}(-\left \vert \alpha \right \vert ^{2})L_{n}(-\left \vert
\alpha \right \vert ^{2})+e^{-4\left \vert \alpha \right \vert ^{2}}L_{m}%
(\left \vert \alpha \right \vert ^{2})L_{n}(\left \vert \alpha \right \vert ^{2}%
)}\nonumber \\
& \times \sqrt{L_{n}^{2}(-\left \vert \alpha \right \vert ^{2})-e^{-4\left \vert
\alpha \right \vert ^{2}}L_{n}^{2}(\left \vert \alpha \right \vert ^{2}%
)},\label{ht34}%
\end{align}
which exhibits the exchanging symmetry with respect to $m$ and $n$, namely,
$C_{+}(\alpha,n,m)=C_{+}(\alpha,m,n).$ In addition, when $m=n$, Eq.(\ref{ht34}%
) becomes
\begin{equation}
C_{+}=\frac{L_{m}^{2}(-\left \vert \alpha \right \vert ^{2})-e^{-4\left \vert
\alpha \right \vert ^{2}}L_{m}^{2}(\left \vert \alpha \right \vert ^{2})}{L_{m}%
^{2}(-\left \vert \alpha \right \vert ^{2})+e^{-4\left \vert \alpha \right \vert
^{2}}L_{m}^{2}(\left \vert \alpha \right \vert ^{2})},\label{ht35}%
\end{equation}
which implies that the two-mode symmetrically excited EECSs is also not a
maximally entangled state. As expected, the concurrence of single-mode excited
EECSs is just a special case of Eq.(\ref{ht34}) \cite{14}. For the two-mode
subtraction ($t_{A}=t_{B}=1$ and $r_{A}=r_{B}=0$), the concurrence $C_{+}$
keep unchanged for the total even number photon-subtraction case ($m+n=$even
number); while for the total odd number photon-subtraction case ($m+n=$odd
number), the states shall present a transformation from the partially
entangled state to the maximum entangled state.

In order to see clearly the effect of different parameters $r_{j},t_{j}$ and
$(m,n)$ on the concurrence of CS-EECSs, $C_{+}$ for the state $\left \vert
\Psi_{+}\left(  \alpha,m,n\right)  \right \rangle $ as a function of different
parameters are shown in Figs.2-3. We show the concurrence $C_{+}$ with a
symmetric case of $r_{A}=r_{B}=r$ and $m=n=1,2$ in Fig.2, from which we can
see that the concurrence $C_{+}$ increases with $r$ ($\alpha$) when $\alpha$
($r$) exceeds a threshold; while the case is still true for $m=n=2$ but
without the threshold. Here, for convenience, we plotted only with $\alpha$
being real number.\begin{figure}[ptb]
\label{Fig6} \centering \includegraphics[width=7cm]{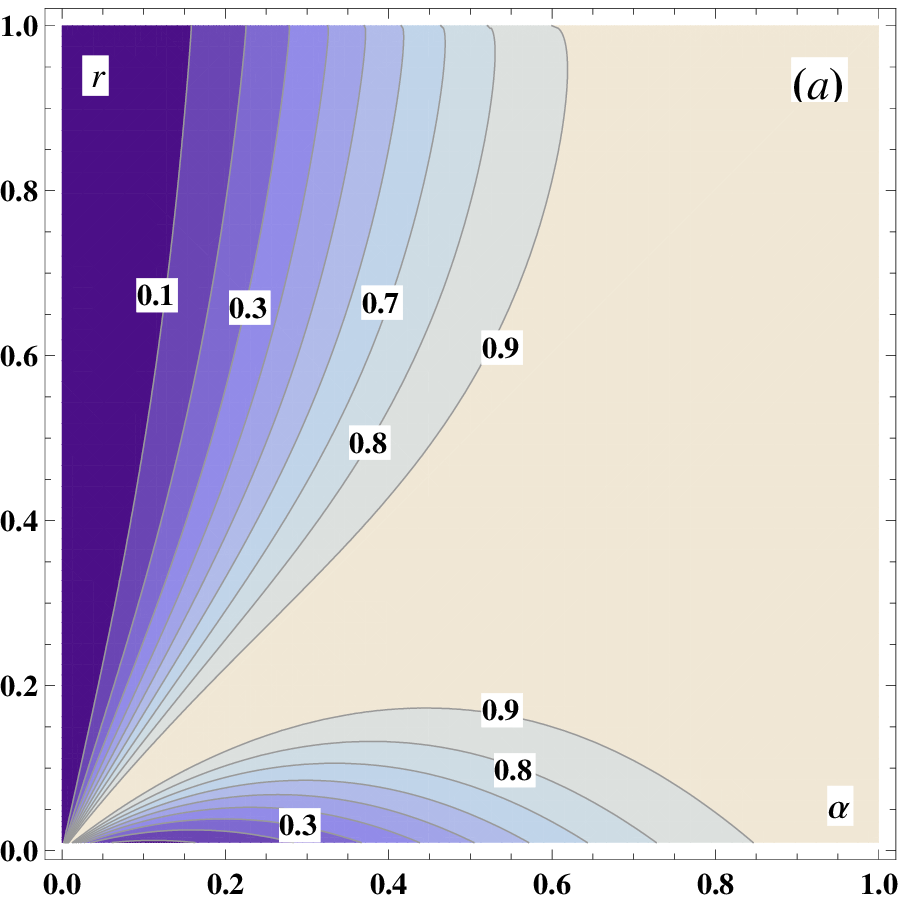}
\  \  \  \includegraphics[width=7cm]{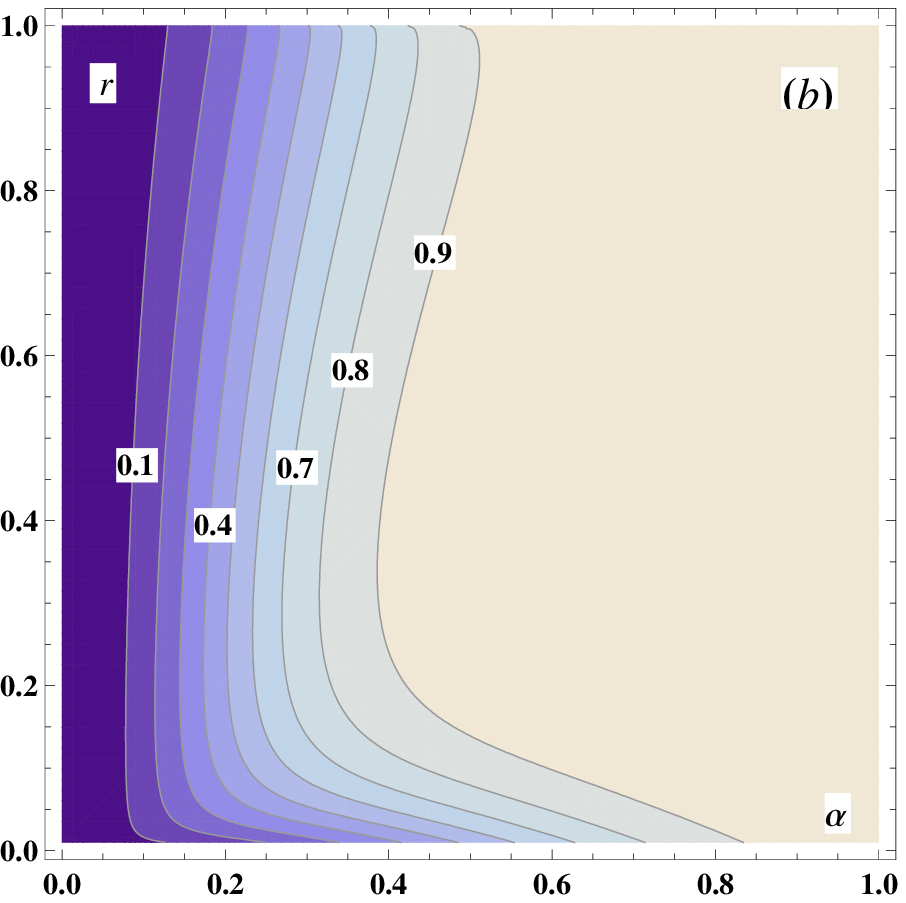}\caption{{}(Color online)
Concurrence of entanglement of $\left \vert \Psi_{+}\left(  \alpha,m,n\right)
\right \rangle $ as a function of $\alpha$ (considered as a real number) and
$r$ for different $m$ and $n$ values. (a) $m=n=1$; (b) $m=n=2$.}%
\end{figure}

From Fig.3(a)-(b) with a given value of $r=1/\sqrt{2}$, one can see that
$C_{+}$ increases with the increase of $\alpha$ for given parameters $m$ and
$n$. Especially, the concurrence $C_{+}$ tends to unit for the larger $\alpha
$. In addition, for a symmetric case of $m=n$, $C_{+}$ increases with the
increase of $m$ (Fig.3(a)); while for an asymmetric case of $m\neq n$, $C_{+}$
increases with the increase of $n$ (Fig.3(b)). Thus the (high-order) coherent
superposition operation can be applied to enhance the entanglement of the
state $\left \vert \Psi_{+}\left(  \alpha,m,n\right)  \right \rangle
.$\begin{figure}[ptb]
\label{Fig7}
\centering \includegraphics[width=7cm]{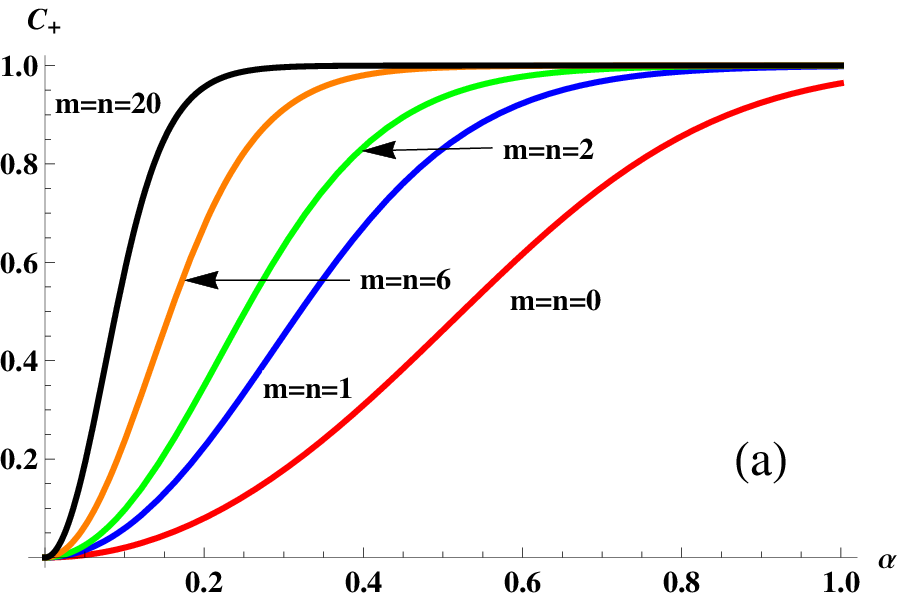}\  \  \includegraphics[width=7cm]{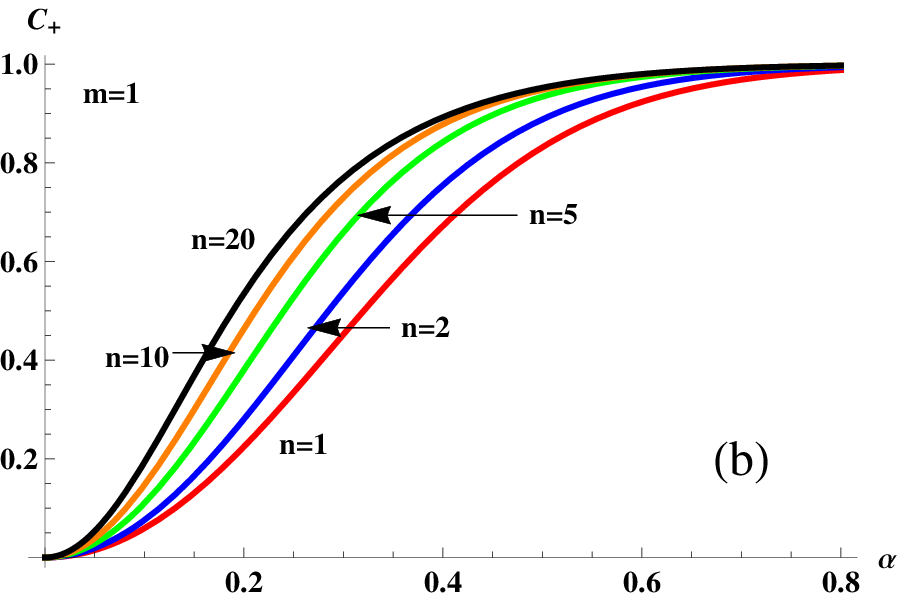}\newline%
\caption{{}(Color online) Concurrence of entanglement for the CS-EECSs
$\left \vert \Psi_{+}\left(  \alpha,m,n\right)  \right \rangle $ with
$r=1/\sqrt{2}$. (a) $m=n$; (b) $m\neq n$.}%
\end{figure}

In particular, we make a comparison of the entanglement properties between the
CS-EECSs (single-mode CS-EECSs $\left \vert \Psi_{+}\left(  \alpha,1,0\right)
\right \rangle $ and $\left \vert \Psi_{+}\left(  \alpha,1,1\right)
\right \rangle )$ and the single(two-)--mode photon excited EECSs ($a^{\dagger
}\left \vert \Psi_{+}\left(  \alpha,0,0\right)  \right \rangle $ and
$a^{\dagger}b^{\dagger}\left \vert \Psi_{+}\left(  \alpha,0,0\right)
\right \rangle $) (see Fig.4). From Fig.4 one can clearly see that: (i) both
photon-excitation and CS operations can improve the entanglement of
$\left \vert \Psi_{+}\left(  \alpha,0,0\right)  \right \rangle ;$ (ii) for
single-mode operations ($ta+ra^{\dagger}$ and $a^{\dagger}$), the effects of
the CS operations are more prominent than those of the mere photon-addition in
a small-amplitude regime. The threshold value is about $\alpha \lesssim0.6$
with a symmetrical case of $r=1/\sqrt{2}$. In addition, the improvement of
entanglement is more obvious by an asymmetrical parameter (say $r=0.4$) than
that by $r=1/\sqrt{2}$. These results show that applying a CS operator on the
EECSs may be better to enhance the quantum entanglement than using only a
creation operator on the EECSs, especially for an asymmetrical parameter $r$.
For two-mode operations ($\left(  ta+ra^{\dagger}\right)  \left(
tb+rb^{\dagger}\right)  $ and $a^{\dagger}b^{\dagger}$), the case is true
(However, we should point that the most optimal case is not for $r=0$ or $t=1$
because two-mode photon-subtraction EECS $ab\left \vert \Psi_{+}\left(
\alpha,0,0\right)  \right \rangle $ is $\left \vert \Psi_{+}\left(
\alpha,0,0\right)  \right \rangle $). (iii) Moreover, the two-mode operations
are more effective on increasing the entanglement of EECSs than the
single-mode operations. The entanglement can be modulated by different values
of $r$, which can also be clearly seen from Fig.3(a).

\begin{figure}[ptb]
\label{Fig71} \centering \includegraphics[width=7cm]{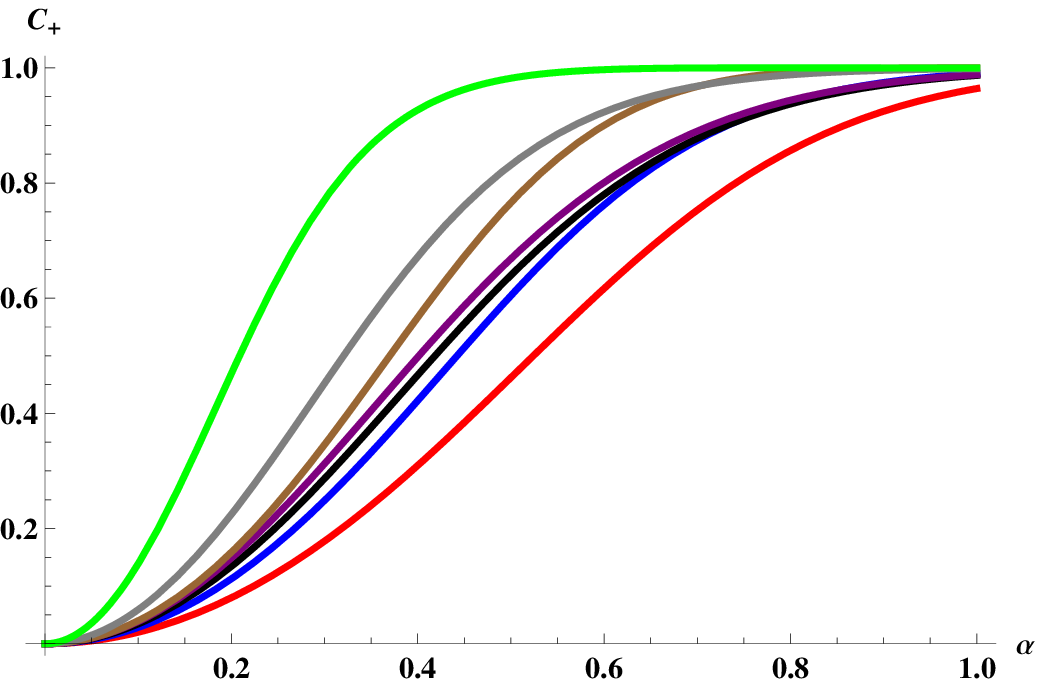}\caption{{}%
(Color online) Concurrence of entanglement for the states: ECSs $\left \vert
\Psi_{+}\left(  \alpha,0,0\right)  \right \rangle $ (Red line), single-photon
excited ECSs $a^{\dagger}\left \vert \Psi_{+}\left(  \alpha,0,0\right)
\right \rangle $ (Blue line), single-mode CS-ECSs $\left \vert \Psi_{+}\left(
\alpha,1,0\right)  \right \rangle $ (Black and Purple lines with $r=1/\sqrt
{2},0.6$ respectively), two-mode excited CESs $a^{\dagger}b^{\dagger
}\left \vert \Psi_{+}\left(  \alpha,0,0\right)  \right \rangle $ (Origin line),
two-mode CS-CESs $\left \vert \Psi_{+}\left(  \alpha,1,1\right)  \right \rangle
$ (Gray and Green lines with $r=1/\sqrt{2},0.4$ respectively).}%
\end{figure}

\section{Fidelity of quantum teleportation}

It is well known that photon subtraction from or photon addition to bipartite
Gaussian states can be used to improve the entanglement and quantum
teleportation \cite{3,4,11,25,26,27,28,29,30,31,32}. In this section, we
consider the effect of coherent superposition on quantum teleportation (QT) by
using the CS-EECSs as entangled resource. In order to describe the quality of
a QT scheme, the fidelity between a unknown input state and the teleported
output state is usually used as a measure. For a continuous-variable (CV)
system, a teleportation scheme has been proposed according to the
characteristic functions (CFs) of the quantum states concluding input, source
and teleported states \cite{33}.

For a two-mode system $\rho$, the CF is defined as $\chi \left(  \alpha
,\beta \right)  =\mathtt{tr}\left[  D_{a}\left(  \eta \right)  D_{b}\left(
\gamma \right)  \rho \right]  $, where $D_{a,b}$ are the displacement operators.
Noticing the displacement operator $D_{a}\left(  \eta \right)  =e^{-\left \vert
\eta \right \vert ^{2}/2}e^{\eta a^{\dagger}}e^{-\eta^{\ast}a}$ and using
(\ref{ht3}), then the CF of the CS-EECSs can be calculated as
\begin{align}
\chi_{E}\left(  \eta,\gamma \right)   &  =N_{\pm,m,n}^{2}\left \{
CF_{A}^{\alpha,\alpha}\left(  \eta \right)  CF_{B}^{\alpha,\alpha}\left(
\gamma \right)  \right.  \nonumber \\
&  +CF_{A}^{-\alpha,-\alpha}\left(  \eta \right)  CF_{B}^{-\alpha,-\alpha
}\left(  \gamma \right)  \nonumber \\
&  \pm e^{-4\left \vert \alpha \right \vert ^{2}}CF_{A}^{\alpha,-\alpha}\left(
\eta \right)  CF_{B}^{\alpha,-\alpha}\left(  \gamma \right)  \nonumber \\
&  \pm \left.  e^{-4\left \vert \alpha \right \vert ^{2}}CF_{A}^{-\alpha,\alpha
}\left(  \eta \right)  CF_{B}^{-\alpha,\alpha}\left(  \gamma \right)  \right \}
,\label{ht36}%
\end{align}
where we have set (noticing $(\beta,\alpha)=(\pm \alpha,\pm \alpha)$)
\begin{align}
&  CF_{j}^{\beta,\alpha}\left(  \eta \right)  \nonumber \\
&  =e^{-\left \vert \eta \right \vert ^{2}/2+\eta \beta^{\ast}-\eta^{\ast}\alpha
}\frac{\partial^{2m}}{\partial \tau^{m}\partial s^{m}}e^{\frac{1}{2}(\tau
^{2}+s^{2})t_{j}r_{j}}\nonumber \\
&  \times \left.  e^{\left(  \eta r_{j}+r_{j}\alpha+\beta^{\ast}t_{j}\right)
\tau+\left(  \alpha t_{j}-\eta^{\ast}r_{j}+r_{j}\beta^{\ast}\right)  s+\tau
sr_{j}^{2}}\right \vert _{s,\tau=0},\label{ht37}\\
&  \left(  j=A,B\right)  .\nonumber
\end{align}
For convenience of further calculation, here we keep the differential form of
$CF_{j}^{\beta,\alpha}\left(  \eta \right)  $ in Eq.(\ref{ht37}).

Next, we consider the Braunstein and Kimble protocol \cite{34} of QT for
single-mode coherent-input states $\left \vert \gamma \right \rangle $. Note that
the fidelity is independent of amplitude of the coherent state, thus for
simplicity we take $\gamma=0$, then we have only to calculate the fidelity of
the vacuum input state with the CF $\chi_{in}\left(  z\right)  =\exp
[-\left \vert z\right \vert ^{2}/2]$. It is shown that, with the CF $\chi
_{E}\left(  \eta,\gamma \right)  $ for entangled channel, the CF $\chi
_{out}\left(  z\right)  $ of the output state can be related to the CFs of
input state and entangled source by formula $\chi_{out}\left(  z\right)
=\chi_{in}\left(  z\right)  \chi_{E}\left(  z^{\ast},z\right)  ,$ and the
fidelity of QT of CVs can be obtained as \cite{33}
\begin{equation}
\mathcal{F=}\int \frac{d^{2}z}{\pi}\chi_{in}\left(  z\right)  \chi_{out}\left(
-z\right)  . \label{ht38}%
\end{equation}

Thus substituting Eqs.(\ref{ht36})-(\ref{ht37}) into Eq. (\ref{ht38}) yields%
\begin{align}
\mathcal{F}_{m,n}^{+}  & =N_{+,m,n}^{2}\left[  \mathcal{F}^{\alpha,\alpha
}+\mathcal{F}^{-\alpha,-\alpha}\right.  \nonumber \\
& +\left.  e^{-4\left \vert \alpha \right \vert ^{2}}\left(  \mathcal{F}%
^{\alpha,-\alpha}+\mathcal{F}^{-\alpha,\alpha}\right)  \right]  ,\label{ht39}%
\end{align}
where we have derived
\begin{align}
&  \mathcal{F}^{\beta,\alpha}\nonumber \\
&  =\frac{1}{2}e^{\frac{1}{2}(\beta^{\ast}-\alpha)^{2}}\sum_{l=0}^{m}%
\sum_{f=0}^{n}\sum_{k,j=0}^{\min(n-f,m-l)}\nonumber \\
&  \times \frac{\left(  m!\right)  ^{2}\left(  n!\right)  ^{2}}{2^{m+n}%
l!f!k!j!}\frac{\left(  -1\right)  ^{k+j}t_{A}^{m-l}t_{B}^{n-f}}{\left(
m-l-k\right)  !\left(  m-l-j\right)  !}\nonumber \\
&  \times \frac{r_{A}^{m+l}r_{B}^{n+f}\left(  r_{A}r_{B}\right)  ^{k+j}\left(
\sqrt{t_{A}r_{A}t_{B}r_{B}}\right)  ^{-k-j}}{\left(  n-f-k\right)  !\left(
n-f-j\right)  !}\nonumber \\
&  \times H_{m-l-k}(N_{2})H_{m-l-j}(N_{1})H_{n-f-k}(M_{2})H_{n-f-j}%
(M_{1}),\label{ht40}%
\end{align}
and set
\begin{align}
N_{1} &  =[\beta^{\ast}t_{A}+\frac{1}{2}r_{A}\left(  \beta^{\ast}%
+\alpha \right)  ]/(-i\sqrt{2t_{A}r_{A}}),\nonumber \\
N_{2} &  =[\alpha t_{A}+\frac{1}{2}r_{A}\left(  \beta^{\ast}+\alpha \right)
]/(i\sqrt{2t_{A}r_{A}}),\nonumber \\
M_{1} &  =[\beta^{\ast}t_{B}+\frac{1}{2}r_{B}\left(  \beta^{\ast}%
+\alpha \right)  ]/(-i\sqrt{2t_{B}r_{B}}),\nonumber \\
M_{2} &  =[\alpha t_{B}+\frac{1}{2}r_{B}\left(  \beta^{\ast}+\alpha \right)
]/(i\sqrt{2t_{B}r_{B}}).\label{ht41}%
\end{align}
Eq.(\ref{ht39}) is just the fidelity of teleporting the coherent state by
using the CS-EECSs as entangled resource. In particular, when $m=n=0$,
Eq.(\ref{ht40}) reduces to $\mathcal{F}^{\beta,\alpha}=\frac{1}{2}e^{\frac
{1}{2}(\beta^{\ast}-\alpha)^{2}}$, submitting it into Eq.(\ref{ht39}) yields%
\begin{equation}
\mathcal{F}_{0,0}^{+}=\frac{e^{\frac{1}{2}(\alpha^{\ast}-\alpha)^{2}%
}+e^{-4\left \vert \alpha \right \vert ^{2}}e^{\frac{1}{2}(\alpha^{\ast}%
+\alpha)^{2}}}{2(1+e^{-4\left \vert \alpha \right \vert ^{2}})},\label{ht42}%
\end{equation}
which is just the fidelity by using the usual EECSs as entangled channel. It
depends on the real and imagine parts of coherent amplitude $\alpha$\ of the
quantum channel, which is different from the case \cite{35,36}. If the
fidelity exceeds the classical limit $1/2$, then the teleportation can be
considered as a successful quantum protocol. From Eq.(\ref{ht42}) it is found
that under a small region of $\alpha$ ($\left \vert \alpha \right \vert <2.2$),
$\mathcal{F}_{0,0}^{+}$ may be larger than $1/2$ (see Fig.5) (the fidelity is
always less than$\ 1/2$ for odd entangled coherent state). This implies that
although the odd entangled coherent state is the maximum entangled coherent
state, the fidelity teleporting coherent state has not been enhanced, and that
the non-maximum entangled state can be used as more effective resource for
quantum teleportation \cite{28,29,30}.

\begin{figure}[h]
\label{Fig2}
\centering \includegraphics[width=8cm]{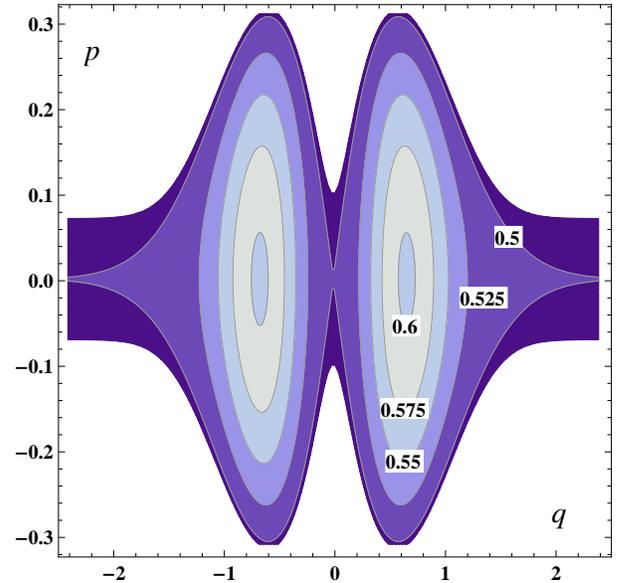}\caption{{\protect \small (Color
online) The fidelity }$\mathcal{F}_{0,0}^{+}${\protect \small of teleporting
coherent state by using ECSs as entangled channel, where }$\alpha
=q+ip${\protect \small . }}%
\end{figure}

In order to know how the CS-EECSs can improve the fidelity of QT, we plot the
fidelity of teleporting coherent state by using CS-EECSs as the entangled
resource with a symmetric case of $r_{A}=r_{B}=r$ $\left(  0<r<1\right)  $ in
Fig.6(a). It is shown that the maximum fidelity for $\mathcal{F}_{1,1}^{+}$
can be realized by the CS operation in the small region of coherent amplitude
$\alpha$ (taking a real number) and the value of $\theta$ for optimal fidelity
will be modulated by different values of small $\alpha$. Generally, the
fidelity increases with the increase of the value of $\theta$. To obtain an
effective QT, i.e., $\mathcal{F}_{1,1}^{+}>1/2$, $r$ must be less than
$r=0.6$. For instance, from Fig.6(a) we can see that the fidelity 0.65 can be
obtained well above the classical limit by the CS, say for $r=0.05$ and
$\alpha=0.1$. This case is true for coherent superposition two-mode squeezed
vacuum \cite{11}. In addition, we should mention that the state $\left \vert
\Psi_{+}\left(  \alpha,1,1\right)  \right \rangle $ becomes $\left \vert
\Psi_{+}\left(  \alpha,1,1\right)  \right \rangle \sim ab\left \vert \Psi
_{+}\left(  \alpha,0,0\right)  \right \rangle \sim \left \vert \Psi_{+}\left(
\alpha,0,0\right)  \right \rangle $ at $r=0$, thus the line of $r=0$ is just
the fidelity by using the EECSs as entangled channel. From this point, we can
see that for a given small value of $\alpha$ the fidelity can be optimized
over the classical limit and over the fidelity at $r=0$. For the case of the
odd entangled coherent state, there is no any effective fidelity found for any
values of $r$ and $\alpha$.

\begin{figure}[h]
\label{Fig5}
\centering \includegraphics[width=7cm]{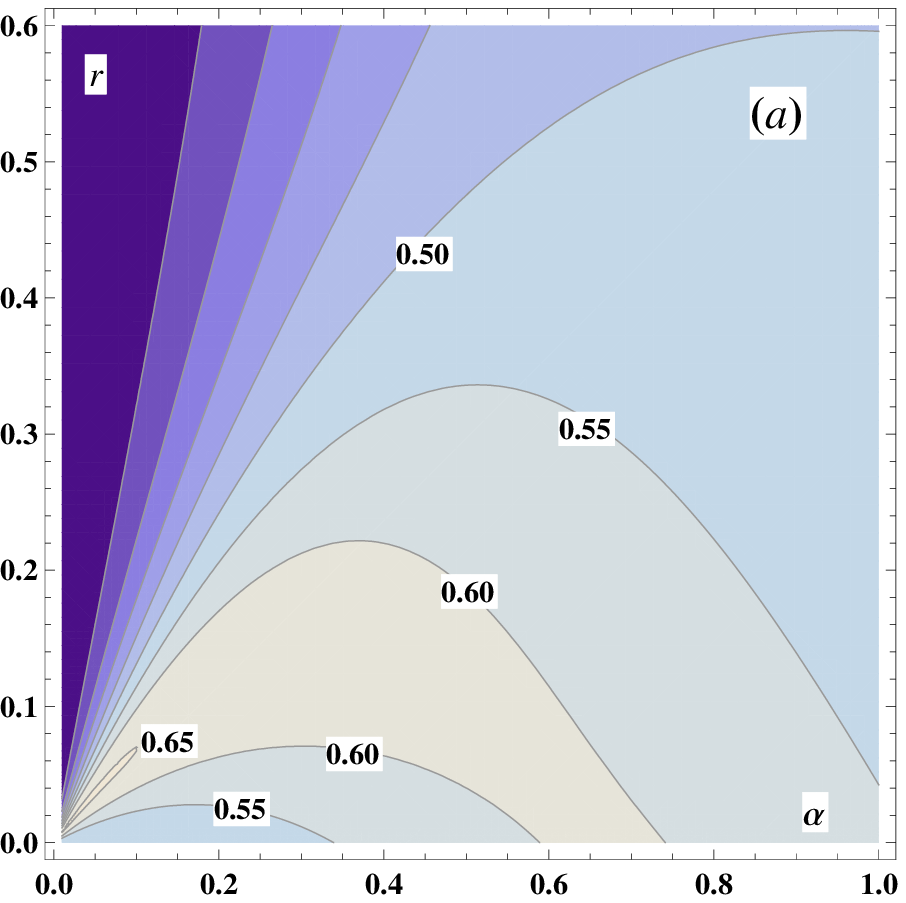}\\ \includegraphics[width=7cm]{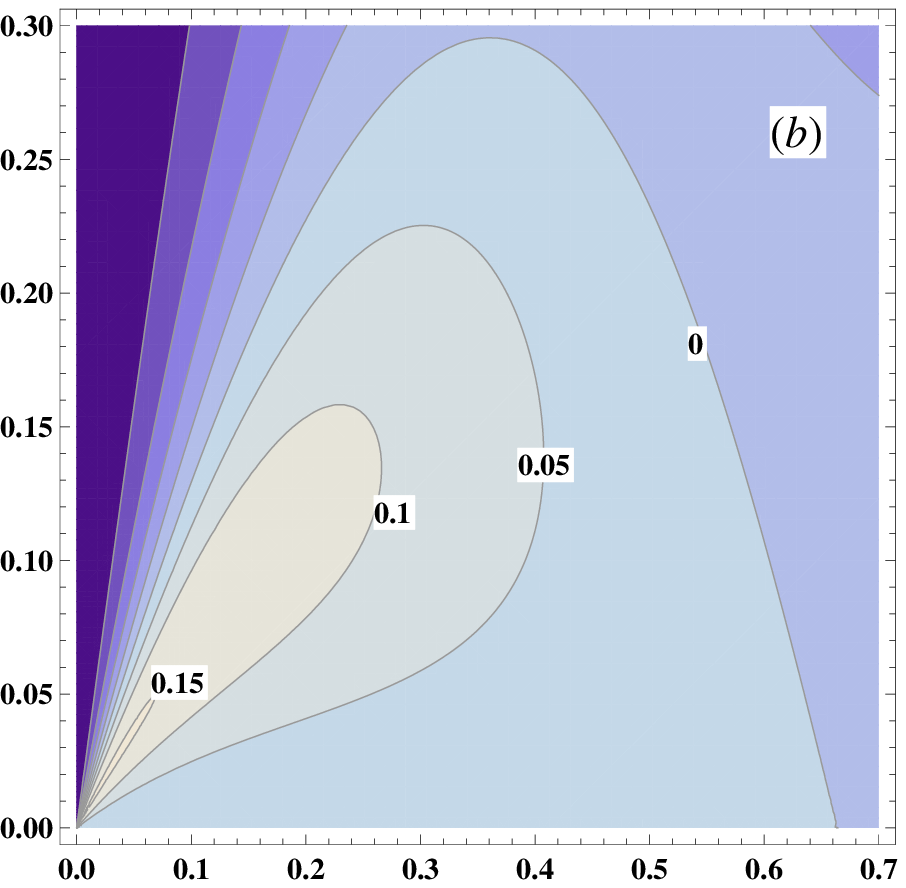}\caption{{\protect \small (Color
online) The fidelity of teleporting coherent state by using CS-EECSs as
entangled channel, where }$\alpha${\protect \small is a real number and }%
$r_{A}=r_{B}=r${\protect \small . (a) }$F_{1,1}^{+}${\protect \small as a
function }$\alpha${\protect \small and }$r${\protect \small ; (b) }$F_{1,1}%
^{+}-F_{0,0}^{+}${\protect \small as a function }$\alpha$ {\protect \small and
}$r${\protect \small ;}}%
\end{figure}

As a comparison between the CS-EECSs and the EECSs, in Fig.6(b) we display the
difference of fidelity ($\Delta \mathcal{F}^{+}=\mathcal{F}_{1,1}%
^{+}-\mathcal{F}_{0,0}^{+}$) as the function of $\alpha$ and $r$. It is shown
that there is an over-zero region for the fidelity difference $\Delta
\mathcal{F}^{+}$ which means that the CS operation can be used to improve the
fidelity teleporting coherent state.

In Fig. 7, we plot the fidelity as a function of $\alpha$ for several
different values of $(m,n)$ with a symmetric case of $m=n$ and $r=\allowbreak
0.195$. Form Fig.7 we can see that the fidelity $\mathcal{F}_{m,m}^{+}$
decrease with the values of $m$ only when $\alpha$ exceeds a certain threshold
value (Fig.7(a)). In particular, $\mathcal{F}_{2,2}^{+}$ may achieve more
optimal value over classical limit than $\mathcal{F}_{1,1}^{+}$ in a small
region of $\alpha$. On the other hand, for an asymmetric case ($m\neq n$,
$r=\allowbreak0.195$), it is interesting to notice that the fidelity
$\mathcal{F}_{m,n}^{+}$ decreases due to the asymmetric operation (Fig.7(b)).
These indicate that multiple CS operations may not only beat the classical
bound but also even surpass single CS operations on each mode under a certain condition.

\begin{figure}[h]
\label{Fig4}
\centering \includegraphics[width=6cm]{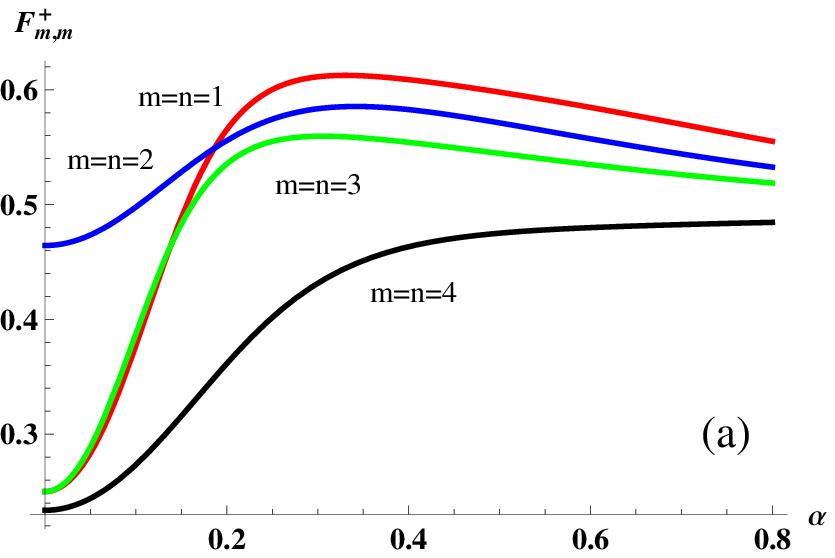}\\ \includegraphics[width=6cm]{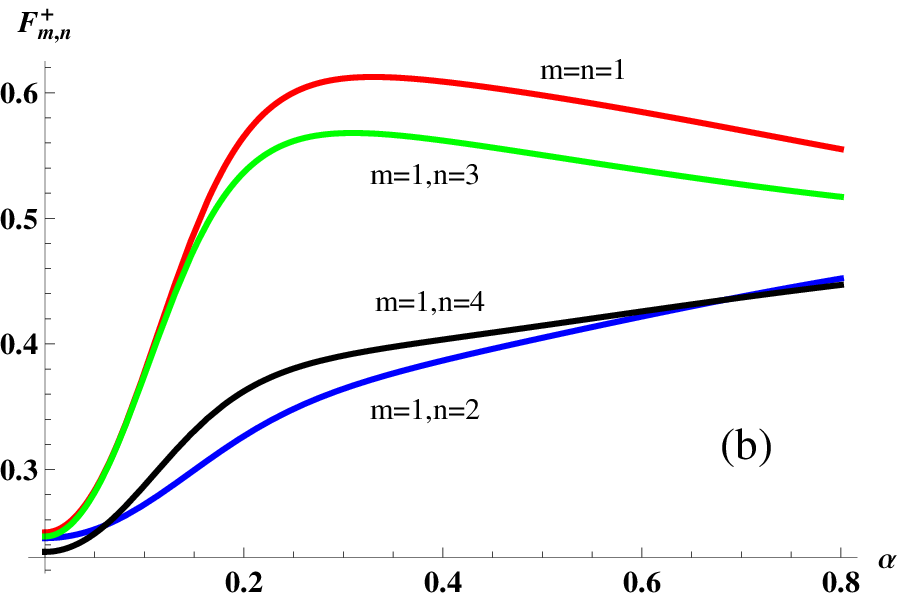}\caption{{\protect \small (Color
online) The fidelity as a function of }$\alpha$ {\protect \small for several
different values of }$(m,n)${\protect \small , where }$\alpha$
{\protect \small is taken a real number and }$r=\allowbreak0.195$%
{\protect \small . }}%
\end{figure}

\section{Conclusions}

In this paper, we have introduced the concept of two-mode coherent
superposition EECSs by coherent operation ($ra^{\dag}+ta$) of photon
subtraction and addition, and investigated the properties of entanglement
according to the concurrence, the Shchukin-Vogel criteria and the average
fidelity of quantum teleportation. It is shown that for the EECS, all of
entanglement characteristics can be improved by local coherent operation. We
made a comparison of the entanglement properties between the CS-EECSs
(single-mode CS-ECSs $\left \vert \Psi_{+}\left(  \alpha,1,0\right)
\right \rangle $ and $\left \vert \Psi_{+}\left(  \alpha,1,1\right)
\right \rangle )$ and the single(two-)--mode photon excited EECSs ($a^{\dagger
}\left \vert \Psi_{+}\left(  \alpha,0,0\right)  \right \rangle $ and
$a^{\dagger}b^{\dagger}\left \vert \Psi_{+}\left(  \alpha,0,0\right)
\right \rangle $), which shows that the effects of improvement by coherent
superposition operation $\left(  ta+ra^{\dagger}\right)  \left(
tb+rb^{\dagger}\right)  $\ are more prominent than those by single
($a^{\dagger}$) and two-photon ($a^{\dagger}b^{\dagger}$) addition under a
small region of amplitude, especially for an asymmetrical parameter $r$. Using
the CS-EECSs with a symmetric case ($r_{A}=r_{B}=r$) as an entangled channel,
the fidelity of teleporting a coherent state is also considered, which
presents that more effective quantum teleportation can be realized by
$\left \vert \Psi_{+}\left(  \alpha,1,1\right)  \right \rangle $ than
$\left \vert \Psi_{+}\left(  \alpha,0,0\right)  \right \rangle $. Although the
odd entangled coherent state is a maximum entangled one, the fidelity of
quantum teleportation by using it as an entangled channel is always smaller
than the classical limit value (1/2). Thus it is implied that a maximum
entangled state may be not appreciate for the effective quantum teleportation.

\textbf{Acknowledgments: } Project supported by the National Natural Science
Foundation of China (Grant Nos. 11264018 and 11164009), the Natural Science
Foundation of Jiangxi Province of China (Grant No. 20132BAB212006), and the
Research Foundation of the Education Department of Jiangxi Province of China
(no GJJ14274) as well as Degree and postgraduate education teaching reform
project of jiangxi province(No. JXYJG-2013-027).

\end{document}